# Charged anisotropic white dwarfs in $f(R, T)$ gravity

Zhe Feng
*CAS Key Laboratory of Theoretical Physics,*
*Institute of Theoretical Physics, Chinese Academy of Sciences,*
*Beijing 100190, P. R. China*
*School of Physical Sciences,*
*University of Chinese Academy of Sciences,*
*Beijing 100049, P. R. China*
*Hohai University, Nanjing 211100, P. R. China*
*fengzhe@itp.ac.cn*

Received 4 August 2024
Revised 30 November 2024
Accepted 1 December 2024
Published 31 January 2025

In the framework of $f(R, T) = R + 2\beta T$ gravity, wherein $R$ denotes the Ricci scalar and $T$ represents the trace of the energy–momentum tensor, the equilibrium structure of charged anisotropic white dwarfs is investigated. The stellar equations for the general scenario are derived, and numerical solutions are obtained for the Chandrasekhar equation of state alongside the charge density distribution proportional to the energy density, $\rho_{\rm ch} = \alpha\rho$. By scrutinizing the properties of the solutions under various parameter combinations, we conducted an exhaustive analysis of the influences of electric charge, modified gravity, and anisotropy on the structural characteristics of white dwarfs. Crucially, white dwarfs can transcend the trivial instances of General Relativity in various ways, thereby demonstrating super-Chandrasekhar phenomena. As an inquiry into the structure of white dwarfs, the findings of this research may contribute to elucidating astronomical observations such as superluminous type Ia supernovae.

## 1. Introduction

General Relativity has withstood an extensive array of observational and experimental validations encompassing both weak gravitational regimes (e.g. Mercury's perihelion precession, gravitational lensing, and gravitational redshift within the solar system) and strong gravitational fields (e.g. pulsar binary systems), thereby establishing itself as a profoundly elegant and successful gravitational theory. Nonetheless, the advent of discoveries such as Type Ia supernovae,[1] the cosmic microwave background radiation,[2,3] and most pivotally, the accelerating expansion of the universe, have progressively unveiled the limitations of Einstein's theory. Even within the domain of pure academic curiosity, there exists a compelling impetus to venture beyond the conventional General Relativity paradigm.[4] Consequently, a multitude of alternative models has been posited.[5]

The most elementary correction method involves substituting the Einstein–Hilbert action with a general function of the Ricci scalar $R$, yielding $f(R)$ gravity,[6–9] wherein $R$ is considered a superfluous degree of freedom. As an illustrious instance, the Starobinsky theory[10] addresses numerous stellar queries. The $f(R) = R^{1+\varepsilon}$ gravity serves as a modification proximate to General Relativity in examining compact astronomical bodies.[11,12] Furthermore, incorporating the trace $T = g_{ab}T^{ab}$ of the matter–energy tensor engenders the $f(R,T)$ theory,[13] which delves deeper into the role of matter within the gravitational field. The $f(R,T)$ theory holds particular relevance for the static equilibrium configuration of compact stars, as illustrated in Refs. 14–22. This area has been vigorously investigated in recent years, and this study follows a parallel trajectory.

White dwarfs, quark stars, neutron stars, and similar entities constitute the compact astronomical objects of significant interest. A white dwarf represents the residual stellar core, predominantly consisting of electron-degenerate matter. It is widely postulated that white dwarfs signify the terminal evolutionary phase of stars whose mass is insufficient to culminate into a neutron star or black hole. This category encompasses over 97% of the stellar population within the Milky Way. As an ancient subject of celestial inquiry, white dwarfs have perpetually captivated the astronomical community's interest in recent years. From a theoretical standpoint, scholars have delved into the magnetized white dwarfs,[23–32] the relationship between white dwarfs and black holes,[33–35] and the enigmatic connections with dark matter.[36–39] Concurrently, on the observational front, astronomers are primarily concerned with isolating and confirming white dwarf samples[31,40–46] from extensive datasets (e.g. Gaia, LISA, LSST), followed by in-depth investigations into their composition and evolutionary trajectories.[34,38,47–52] Meanwhile, the physics community has increasingly focused its attention on the detection of gravitational waves[53–55] associated with white dwarfs. By juxtaposing results[49] with those derived from Newtonian gravity, General Relativity is indispensable for determining the properties of white dwarfs and other compact stellar objects, such as the mass–radius relationship and surface gravity, among others.[47,52,56–58] Furthermore, Chandrasekhar delineated a stringent mass threshold for white dwarfs,[59] stipulating that their mass cannot surpass $1.44\,M_{\rm sun}$. Should white dwarf masses exceed this boundary, type Ia supernovae detonations may transpire, exhibiting uniform luminosity thus serving as standard candles. Nevertheless, the identification of a subset of ultra-luminous supernovae[60,61] has propounded the conjecture that their progenitors were white dwarfs with super-Chandrasekhar masses.[25] Regrettably, in the absence of direct observational evidence, conjectures pertaining to super-Chandrasekhar white dwarfs remain indirect. In the scope of General Relativity and modified gravity, various possible white dwarf structures with Chandrasekhar equation of state[62] have been extensively studied, including uniformly charged white dwarfs under General Relativity,[63] white dwarfs under $R^2$ gravity,[64] white dwarfs under $f(R,T)$ gravity,[14] charged white dwarfs under General Relativity[65] and charged white dwarfs under $f(R,T)$ gravity,[15] the charge of the latter two are mainly concentrated on the surface of stars. It should be acknowledged that static electricity is rather uncommon in astrophysical observations, with most celestial systems tending toward electrical neutrality to achieve stability.[66] However, Ray *et al.*[67] explored the influence of charge on the structure of compact stars, and subsequent related studies further validated this possibility.[68–70] Consequently, the theoretical investigation of charged white dwarfs and other compact stars retains a certain allure.

The conceptualization of stellar structure elucidated in this discourse primarily emanates from the investigations of numerous scholars concerning quark stars. A quark star constitutes a theoretical variety of compact, exotic stellar objects where the extreme core temperature and pressure compel nuclear particles to form quark matter, a continuous state of matter comprising free quarks. For quark stars, a plethora of structural attributes has been scrutinized,[16–22,71,72] encompassing: surface or internal charge, isotropic or anisotropic pressure, General Relativity or $f(R,T)$ gravity. Conventionally, the MIT Bag Model equation of state is employed when addressing quark star, albeit certain scholars[22] prefer to incorporate the equation of state with $\mathcal{O}(m_s^4)$ correction terms to attain a more precise model. Despite this paper's focus on the structural analysis of white dwarfs, it introduces a charge density proportional to the energy density and an anisotropic pressure distribution.

This paper is structured as follows. Subsequent to this introduction (Sec. 1), Sec. 2 offers a succinct review of $f(R,T)$ gravity and derives the field equations in the presence of matter and electromagnetic fields. In Sec. 3, the modified Tolman–Oppenheimer–Volkoff equations (Sec. 3.1) are derived considering the metric of spherical symmetry. The Chandrasekhar equation of state (Sec. 3.2) and the charge distribution (Sec. 3.3) are elaborated upon in two subsections within Sec. 3. In Sec. 4, we procured the numerical solutions of the equations and performed an exhaustive analysis of the influences of electric charge, modified gravity, and anisotropy on the structure of white dwarfs from multiple perspectives. Finally, Sec. 5 synthesizes the conclusions derived from the results.

## 2. $f(R,T)$ Gravity Formalism

Proposed by Harko *et al.*[13] $f(R,T)$ gravity extends the framework of $f(R)$ gravity.[7,6] In this theory, the gravitational action encompasses an arbitrary function of the Ricci scalar $R$ and the trace $T$ of the energy–momentum tensor. When accounting for the presence of both matter and gravitational fields, consider the

following actions:

$$S = \int d^4x\sqrt{-g}[\mathcal{L}_g + \mathcal{L}_m + \mathcal{L}_e] \\ = \int d^4x\sqrt{-g}\left[\frac{1}{16\pi}f(R,T) + \mathcal{L}_m + \mathcal{L}_e\right], \tag{1}$$

where $g$ denotes the determinant of the spacetime metric $g_{ab}$, and $\mathcal{L}_m$ represents the Lagrangian density of the matter field. $\mathcal{L}_e$ is the Lagrangian density of the electromagnetic field, formulated as follows:

$$\mathcal{L}_e = j^a A_a - \frac{1}{16\pi}F_{ab}F^{ab}, \tag{2}$$

where $j^a = \rho_{\rm ch}u^a$ denotes the four-current density, with $\rho_{\rm ch}$ representing the electric charge density and $u^a$ the four-velocity. $A_a$ is the electromagnetic four-potential, and $F_{ab} = \nabla_a A_b - \nabla_b A_a$ is the electromagnetic field strength tensor. Additionally, $\nabla_a$ is the covariant derivative operator compatible with the metric $g_{ab}$.

The field equations can be derived by varying the action with respect to the metric. Initially, the actions of the electromagnetic field and matter are varied to obtain the electromagnetic energy–momentum tensor and the matter energy–momentum tensor, respectively,

$$\mathcal{E}_{ab} = \frac{-2}{\sqrt{-g}}\frac{\delta S_e}{\delta g^{ab}} = \frac{1}{4\pi}\left[g^{cd}F_{ac}F_{bd} - \frac{1}{4}g_{ab}F_{cd}F^{cd}\right], \tag{3}$$

$$\mathcal{M}_{ab} = \frac{-2}{\sqrt{-g}}\frac{\delta S_m}{\delta g^{ab}} = g_{ab}\mathcal{L}_m - 2\frac{\partial \mathcal{L}_m}{\partial g^{ab}}. \tag{4}$$

The total energy–momentum tensor is the summation of the two, denoted as $T_{ab} = \mathcal{M}ab + \mathcal{E}ab$. Given that the electromagnetic energy–momentum tensor is traceless, we have $T \equiv g^{ab}T_{ab} = g^{ab}\mathcal{M}_{ab} \equiv \mathcal{M}$.

The variation of the action of the gravitational field is continued with a view to obtain the equations of the gravitational field.

$$\frac{16\pi}{\sqrt{-g}}\frac{\delta S_g}{\delta g^{ab}} = f_R R_{ab} - \frac{1}{2}g_{ab}f + (g_{ab}\Box - \nabla_a\nabla_b)f_R + f_T(\mathcal{M}_{ab} + \Theta_{ab}), \tag{5}$$

where $f_R \equiv \partial f(R,T)/\partial R$, $f_T \equiv \partial f(R,T)/\partial T$, $\Box = \nabla^a\nabla_a$ is the d'Alembert operator, and

$$\Theta_{ab} \equiv g^{cd}\frac{\delta \mathcal{M}_{cd}}{\delta g^{ab}} = -2\mathcal{M}_{ab} + g_{ab}\mathcal{L}_m - 2g^{cd}\frac{\partial^2 \mathcal{L}_m}{\partial g^{ab}\partial g^{cd}}. \tag{6}$$

According to the variational principle $\delta S = 0$, we can get the equation of motion (EoM) of the gravitational field

$$f_R R_{ab} - \frac{1}{2}g_{ab}f + (g_{ab}\Box - \nabla_a\nabla_b)f_R = 8\pi(\mathcal{M}_{ab} + \mathcal{E}_{ab}) - f_T(\mathcal{M}_{ab} + \Theta_{ab}). \tag{7}$$

Analogous to the $f(R)$ theory, the Ricci scalar is regarded as an extraneous degree of freedom, whose EoM can be derived by taking the trace of the EoM tensor

$$3\Box f_R + Rf_R - 2f = 8\pi\mathcal{M} - f_T(\mathcal{M} + \Theta). \tag{8}$$

Similar to the case in General Relativity, taking the covariant derivative of the EoM tensor, Eq. (7) yields an equation for the energy–momentum tensor

$$\nabla^a \mathcal{M}_{ab} = \frac{f_T}{8\pi - f_T}\left[(\mathcal{M}_{ab} + \Theta_{ab})\nabla^a \ln f_T + \nabla^a \Theta_{ab} - \frac{1}{2}g_{ab}\nabla^a T - \frac{8\pi}{f_T}\nabla^a \mathcal{E}_{ab}\right]. \tag{9}$$

It can be substantiated that if $f(R,T) = R$ is adopted, the theory reverts to the canonical framework of General Relativity: Eq. (7) reverts to Einstein's gravitational field equations, while Eq. (9) reverts to the conservation of energy–momentum.

Next, we will take $f(R,T) = R + 2\beta T$, following the examples in Refs. 14–22 and 73–75. Herein $\beta$ is dimensionless in the geometrized unit system, which can be conspicuously inferred from Einstein's equation through dimensional analysis.

## 3. Stellar Structure Equations

### 3.1. *Modified Tolman–Oppenheimer–Volkoff equations*

In order to further decompose the equation into a component form, a static spherically symmetric spacetime line element is selected as

$$\mathrm{d}s^2 = -\mathrm{e}^{2\psi(r)}\mathrm{d}t^2 + \mathrm{e}^{2\lambda(r)}\mathrm{d}r^2 + r^2(\mathrm{d}\theta^2 + \sin^2\theta \mathrm{d}\phi^2). \tag{10}$$

At this point, the Maxwell equation obeyed by the electromagnetic field can be written more explicitly

$$\nabla^a F_{ab} = -4\pi j_b, \nabla_{[a}F_{bc]} = 0. \tag{11}$$

It has only two nonzero components, namely

$$F^{01} = -F^{10} = \frac{q(r)}{r^2}\mathrm{e}^{-\psi(r)-\lambda(r)}, \tag{12}$$

where the charge function has the following form:

$$q(r) = 4\pi\int_0^r \bar{r}^2 \rho_{\mathrm{ch}}(\bar{r})\mathrm{e}^{\lambda(\bar{r})}\mathrm{d}\bar{r}. \tag{13}$$

where $\rho_{\mathrm{ch}}$ is the charge density, see Sec. 3.3.

Following the methodologies delineated in Refs. 17–21, we opt for the anisotropic matter energy–momentum density, specifically

$$\mathcal{M}_{ab} = (\rho + p_t)u_a u_b + p_t g_{ab} - \sigma k_a k_b, \tag{14}$$

where $u^a$ denotes the four-velocity of the fluid, satisfying $u_a u^a = -1$, and $k^a$ represents a unit radial four-vector, satisfying $k_a k^a = 1$. Considering the metric equation (10), their nonzero components can be explicitly written as $u^0 = \mathrm{e}^{-\psi}$ and $k^1 = \mathrm{e}^{-\lambda}$. The term $\sigma \equiv p_t - p_r$ quantifies the degree to which the star deviates from isotropy: the discrepancy between the transverse pressure $p_t$ and the radial pressure $p_r$. *A posteriori*, we can now establish the material Lagrangian density as $\mathcal{L}_m = p_r$, a modification of the widely adopted $\mathcal{L}_m = p$ within the literature[18–21,76,77] under anisotropic configuration. Notably, the assertion in Ref. 78 posits that $\mathcal{L}_m = p$ and $\mathcal{L}_m = -\rho$ are consistent only when the material exhibits minimal coupling with the gravitational field. Meanwhile, Harko[79] argued that the material Lagrangian expressed solely in terms of $\rho$ and $p$ is entirely equivalent, given that they are freely interchangeable thermodynamic quantities. Regardless, our choice of $\mathcal{L}_m = p_r$ does not compromise generality, thus ensuring accurate results.

Up to this point, it has been possible to write Eq. (7) in coordinate form, with its three nonzero components as

$$\frac{1}{r^2}\frac{\mathrm{d}}{\mathrm{d}r}(r\mathrm{e}^{-2\lambda}) - \frac{1}{r^2} = -8\pi\left(\rho + \frac{q^2}{8\pi r^4}\right) + \beta(-3\rho + p_r), \tag{15a}$$

$$\frac{1}{\mathrm{e}^{2\lambda}}\left(\frac{2}{r}\psi' + \frac{1}{r^2}\right) = 8\pi\left(p_r - \frac{q^2}{8\pi r^4}\right) + \beta(-\rho + 3p_r), \tag{15b}$$

$$\frac{1}{\mathrm{e}^{2\lambda}}\left(\psi'' + \psi'^2 - \psi'\lambda' + \frac{1}{r}(\psi' - \lambda')\right) = 8\pi\left(p_t + \frac{q^2}{8\pi r^4}\right) + \beta(-\rho + p_r + 2p_t), \tag{15c}$$

Appropriate minus signs have been incorporated to facilitate comparison with the results of other authors. Equation (8) essentially transforms into a calculus concerning the Ricci scalar, which lies beyond the scope of this work. The sole nonzero component of Eq. (9) yields

$$p_r' = -\left(\frac{\rho + p_r}{1 + a}\right)\psi' + \left(\frac{1 - 2a}{1 + a}\right)\frac{\rho_{\mathrm{ch}}\mathrm{e}^{\lambda}q}{r^2} + \left(\frac{a}{1 + a}\right)\rho' + \left(\frac{2}{1 + a}\right)\frac{\sigma}{r}, \tag{16}$$

where $a \equiv \beta/(8\pi + 2\beta)$. For conceptual convenience, following Refs. 12,14,15,22 and 65, a mass function $m(r)$ is introduced and its integral expression is given as

$$m(r) \equiv \int_0^r \left[4\pi r^2\rho + \frac{\beta r^2}{2}(3\rho - p_r) + \frac{qq'}{r}\right]\mathrm{d}r. \tag{17}$$

The first term within the square brackets signifies the contribution of the matter–energy itself, the second term accounts for the correction to the mass function attributable to $f(R, T)$ gravity, and the third term delineates the gravitational impact due to the augmented energy density ensuing from the electrification of the matter. Employing Eqs. (15a) and (17) can be streamlined as follows:

$$\mathrm{e}^{-2\lambda} = 1 - \frac{2m}{r} + \frac{q^2}{r^2}. \tag{18}$$

Indeed, this technique can be perceived as a mathematical device: introducing a new function $m(r)$ while concurrently eliminating the function $\lambda(r)$.

### 3.2. *Chandrasekhar equation of state and anisotropy profile*

White dwarfs can be conceptualized as stellar remnants primarily constituted of a degenerate electron gas intermixed with atomic nuclei. This composition arises from the collapse of a progenitor star, whereby the electron degeneracy pressure counterbalances gravitational forces, thereby stabilizing the structure against further collapse. Simultaneously, the more granular internal constituents are being meticulously scrutinized.[34,38,52] In the study on them, the Chandrasekhar equation of state[62] is usually chosen, written in the parametric form as

$$p_r = \frac{\pi m_{\rm e}^4}{3h^3}\left[x_{\rm F}(2x_{\rm F}^2 - 3)\sqrt{x_{\rm F}^2+1} + 3\sinh^{-1}x_{\rm F}\right], \tag{19}$$

$$\rho = \frac{8\pi\mu_{\rm e} m_{\rm H} m_{\rm e}^3}{3h^3} x_{\rm F}^3, \tag{20}$$

where $x_F \equiv p_F/m_{\rm e}c$ with $p_F$ representing the Fermi momentum, $m_{\rm e}$ the electron mass, $m_{\rm H}$ the nucleon mass, $h$ the Planck constant. $\mu_{\rm e}$ is understood as the ratio between the nucleon number and atomic number for ions. For He, Ca, and O white dwarfs, $\mu_{\rm e} = 2$. For the sake of comprehensiveness, it is imperative to acknowledge that subsequent to Chandrasekhar's completely degenerate relativistic electron gas model with no interaction, attempts to consider additional effects persist. Among these, the magnum opus should be ascribed to Salpeter, whose inventive treatise[80,81] extensively explores numerous corrective effects, encompassing Coulomb interactions, Thomas–Fermi deviation of electron distributions, exchange energy, and spin–spin interactions between the electrons.

The anisotropy of the stellar structure is conceptualized as the disparity between radial and tangential pressures. This approach allows for an augmentation of the stellar structural complexity while preserving spherical symmetry, which remains computationally manageable. In accordance with this framework, the differential between radial and tangential pressures, $\sigma(r)$, is posited as follows:[12,82,83]

$$\sigma = \kappa p_r(1 - {\rm e}^{-2\lambda}). \tag{21}$$

Here, $\kappa$ is dimensionless and $(1 - {\rm e}^{-2\lambda})$ represents a modified version of "compactness" $\mu \equiv 2m(r)/r$ when dealing with charged celestial bodies. The inclusion of "compactness" ensures that anisotropy persists in the limit approaching the stellar core. On the other hand, in the weak gravitational limit, the line element will degenerate into a Minkowski line element, at which point $\sigma(r)$ vanishes. This means that such anisotropy only becomes significant under high-energy density conditions. Conversely, in regions near the stellar surface where the energy density is relatively low, anisotropy ceases to play a significant role, which aligns with astrophysical expectations.

### 3.3. *Charge distribution*

Finally, following Ref. 63, we will assume that the charge density of the white dwarf is proportional to the energy density, i.e.

$$\rho_{\rm ch} = \alpha \rho. \tag{22}$$

Here, $\alpha$ represents the constant proportional coefficient which possesses the dimensionality $[\alpha] = [\rho_{\rm ch}]/[\rho] = [Q]/[M] = [I][T][M]^{-1}$ in the International System of Units (SI), whereas in the geometrized unit system, this relationship simplifies to $[\alpha] = [I]$. It is noteworthy that some authors,[15,65] in their analysis of white dwarfs, propose that the charge density assumes a Gaussian distribution centered on the stellar surface.

## 4. Numerical Results

With all the necessary conditions delineated, the groundwork is set for deriving numerical solutions to the problem. We summarize the requisite equations as follows:

$$\frac{\mathrm{d}q}{\mathrm{d}r} = 4\pi r^2 \rho_{\rm ch} \mathrm{e}^{\lambda}, \tag{23a}$$

$$\frac{\mathrm{d}m}{\mathrm{d}r} = 4\pi r^2 \rho + \frac{\beta r^2}{2}(3\rho - p_r) + \frac{qq'}{r}, \tag{23b}$$

$$\begin{aligned} \frac{\mathrm{d}p_r}{\mathrm{d}r} = &-\left(\frac{\rho + p_r}{1+a}\right)\left[4\pi r p_r + \frac{m}{r^2} - \frac{\beta r}{2}(\rho - 3p_r) - \frac{q^2}{r^3}\right]\mathrm{e}^{2\lambda} \\ &+ \frac{1-2a}{1+a}\frac{q}{4\pi r^4}\frac{\mathrm{d}q}{\mathrm{d}r} + \frac{a}{1+a}\frac{\mathrm{d}\rho}{\mathrm{d}r} + \frac{2}{1+a}\frac{\sigma}{r}. \end{aligned} \tag{23c}$$

Equation (23a) is derived directly from the definition of the charge function in Eq. (13). Similarly, Eq. (23b) follows directly from the definition of the mass function in Eq. (17). The derivation of Eq. (23c), however, is somewhat more intricate. Initially, Eq. (15b) is treated as an algebraic equation to derive $\psi'$.

$$\psi' = \left[4\pi r p_r + \frac{m}{r^2} - \frac{\beta r}{2}(\rho - 3p_r) - \frac{q^2}{r^3}\right]\mathrm{e}^{2\lambda}. \tag{24}$$

The result is derived by substituting it into Eq. (16). The system of equations is characterized by three parameters: $\rho_{\rm ch}$ and $q$ encompass $\alpha$, which signifies the star's electrical properties; $\beta$ (from which $a$ is derived) represents the deviation of $f(R,T)$ theory from General Relativity; and $\kappa$ originates from the anisotropy profile, reflecting the disparity between radial and tangential pressures, thus indicating the deviation from isotropy. To solve the system explicitly, the following central boundary conditions are specified:

$$q(0) = 0, \quad m(0) = 0, \quad \rho(0) = \rho_{\rm c}, \tag{25}$$

where $\rho_{\rm c}$ denotes the matter density at the spherical center of the star. In the quest for numerical solutions, we have elected prototypical values for the central density, specifically Log $\rho_c[\mathrm{kg/m^3}] \in [10, 15]$. Additionally, appropriate parameters $(\alpha, \beta, \kappa)$ are meticulously selected to engender notable but not ostentatious deviations from trivial outcomes, given that extant theories have already elucidated the vast majority of astronomical phenomena comprehensively. Following the practice of the literature, the surface of the star is chosen as the zero point of radial pressure, i.e. $p_r(R_{\mathrm{sur}}) = 0$. Concurrently, the aggregate mass of the star is delineated by the value of the mass function at the surface $M = m(R_{\mathrm{sur}})$. Outside the star, matter no longer exists, but the external electromagnetic field still has a gravitational effect because the star is electrically charged. Externally, $\mathcal{M}_{ab} = 0$ leads to $T = 0$, and in fact $f(R,T) = R + 2\beta T$ theory has degenerated into standard General Relativity theory. Therefore, the external solution has the form of the Reissner–Nordström exterior solution in General Relativity, and the parameter values can be obtained from the continuity at the boundary $R_{\mathrm{sur}}$. The results of the solution are shown in the figures and briefly described as follows. Beyond the stellar boundary, matter ceases to exist, yet the external electromagnetic field continues to exert a gravitational effect due to the star's electric charge. Externally, $\mathcal{M}_{ab} = 0$ implies $T = 0$, effectively reducing the $f(R,T) = R + 2\beta T$ theory to standard General Relativity. Consequently, the external solution adopts the form of the Reissner–Nordström exterior solution in General Relativity. The parameter values are determined by continuity conditions at the boundary $R_{\mathrm{sur}}$. The resultant solutions are illustrated in the accompanying figures and are succinctly described as follows.

Initially, in Fig. 1, we present the commonly scrutinized mass–radius relationship and mass-central density curve. In each illustration, the trivial case $(\alpha = \beta = \kappa = 0)$ is retained for comparative purposes. The three subplots, respectively, elucidate the effects of pure electric charge, pure modified gravity, and pure anisotropy on the stellar radius and total mass. The influence of electric charge manifests as an overall elevation of both curves, while the impacts of modified gravity and anisotropy on the structure of white dwarfs are predominantly concentrated on the low-density and high-density ends, respectively.

Figure 2 delineates the variation of the mass function with respect to radial coordinates across various parameter configurations. A specific central density has been selected to accentuate the deviations of additional effects from the trivial scenario. From the six panels, it is evident that the introduction of electric charge results in a slight increase in the stellar radius while significantly augmenting the total mass. Correspondingly, modified gravity markedly extends the stellar radius with a comparatively smaller increase in mass. Most intriguing is the emergence of anisotropy, which reduces the stellar radius while simultaneously increasing the total mass, thereby enhancing the stellar density.

In Fig. 3, we exhibit the charge distribution function $q(r)$ and the electric field $E(r)$. Initially, the scenario of pure electric charge is depicted, followed by an investigation into the influences of modified gravity and anisotropy on the charge distribution and electric field at fixed parameter $\alpha$. Analogous to the mass function, the charge function monotonically increases, reaching its maximum at the

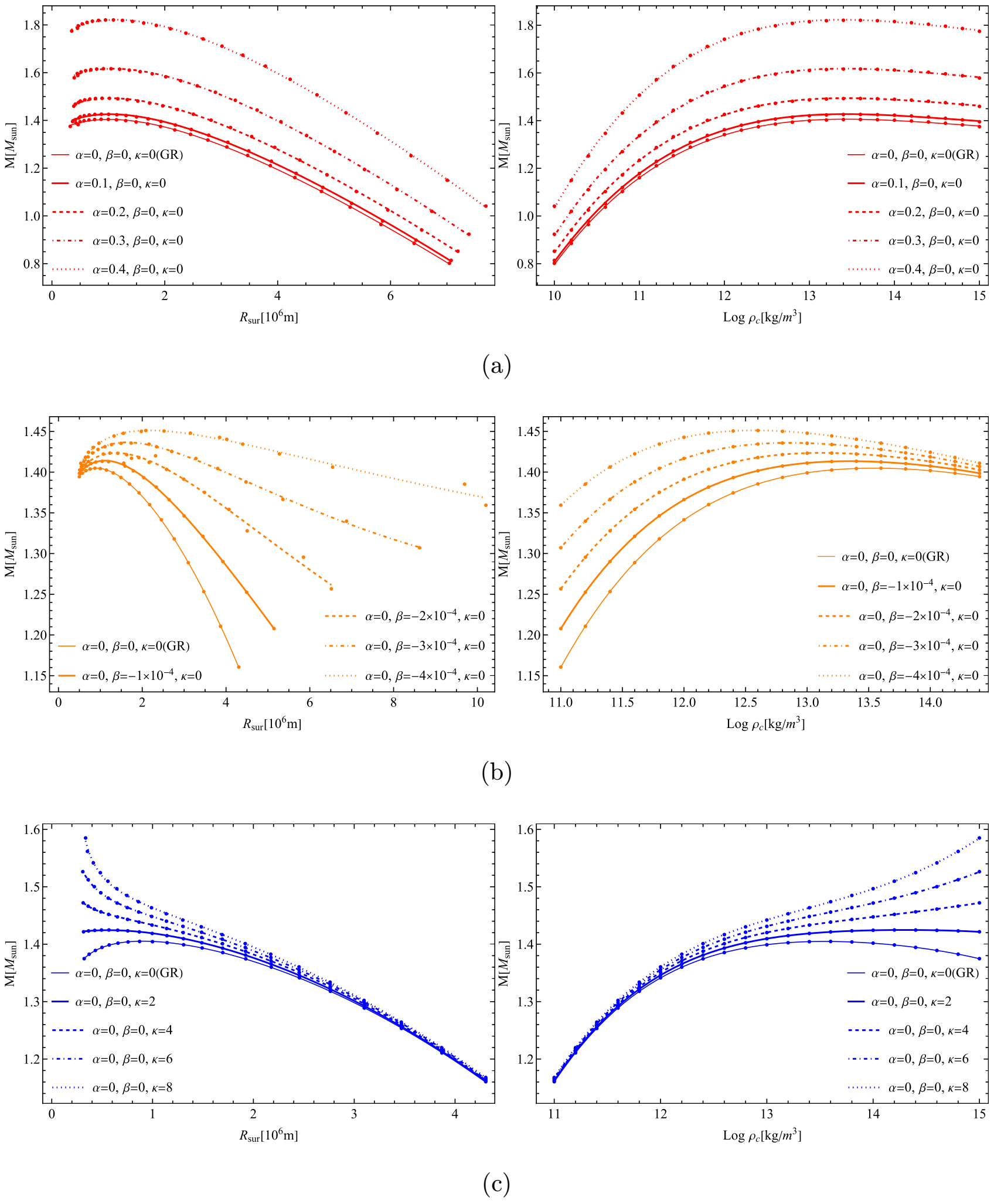

Fig. 1. Mass–radius relation (left column) and mass-central density curve (right column). Notice that the effects of modified gravity and anisotropy exhibit pronounced divergences. (a) Mass–radius relation and mass-central density curve for Log $\rho_c[\mathrm{kg/m^3}] \in [10, 15], \alpha \in [0, 0.4], \beta = 0, \kappa = 0$. We observe that as the ratio of charge density to mass density $\alpha$ increases, the total mass and its upper limit exhibit accelerated growth. (b) Mass–radius relation and mass-central density curve for Log $\rho_c[\mathrm{kg/m^3}] \in [11, 14.5]$, $\alpha = 0, \beta \in [-4 \times 10^{-4}, 0], \kappa = 0$. We observe that as deviations from General Relativity become more pronounced, the values of mass and stellar radius at lower densities exhibit significant variations compared to the trivial case. Conversely, at higher densities, the structure of white dwarfs remains impervious to such modified gravitational effects. The deviations between the data points and the fitted curve in the left plot are attributable to errors inherent in the methods employed for solving the differential equations and the limitations of machine precision. (c) Mass–radius relation and mass-central density curve for Log $\rho_c[\mathrm{kg/m^3}] \in [10, 15], \alpha = 0, \beta = 0, \kappa \in [0, 8]$. In stark contrast to the aforementioned purely modified gravity scenarios, intriguing phenomena under pure anisotropy are predominantly concentrated on the high-density end. Specifically, the curves of total mass versus stellar radius and central density exhibit markedly disparate characteristics. In contrast, the effects of anisotropy are relatively unremarkable on the lower-density end.

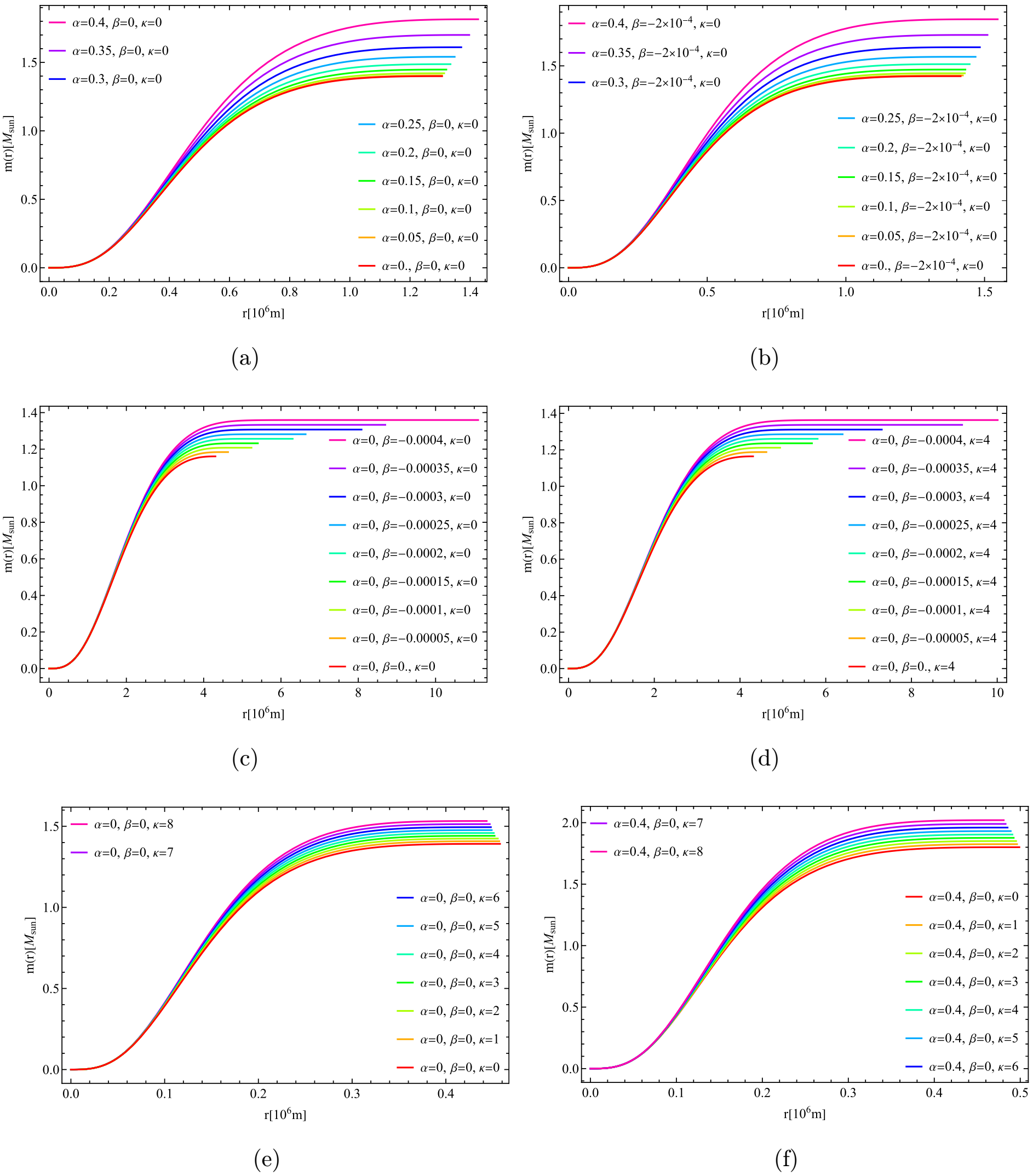

Fig. 2. Mass function $m(r)$. We conducted a comprehensive comparison of the impacts of electric charge, modified gravity, and anisotropy on the structure of white dwarfs, with particular emphasis on the variations in total mass and stellar radius. (a) Mass function for $\rho_c = 10^{13}\,\mathrm{kg/m^3}, \alpha \in [0, 0.4], \beta = 0, \kappa = 0$. The total mass and stellar radius both experience an accelerated augmentation concomitantly with the increase in electric charge. (b) Mass function for $\rho_c = 10^{13}\,\mathrm{kg/m^3}, \alpha \in [0, 0.4], \beta = -2 \times 10^{-4}, \kappa = 0$. Contrasted with the left illustration, the primary effect of modified gravity is to extend the stellar radius while maintaining the total mass. (c) Mass function for $\rho_c = 10^{11}\,\mathrm{kg/m^3}, \alpha = 0, \beta \in [-4 \times 10^{-4}, 0], \kappa = 0$. Consistent with the aforementioned, the influence of modified gravity on the stellar radius surpasses its effect on the total mass. (d) Mass function for $\rho_c = 10^{11}\,\mathrm{kg/m^3}, \alpha = 0, \beta \in [-4 \times 10^{-4}, 0], \kappa = 4$. Due to the comparatively diminutive central density, the effects of anisotropy are not markedly evident at this juncture. (e) Mass function for $\rho_c = 10^{14.5}\,\mathrm{kg/m^3}, \alpha = 0, \beta = 0, \kappa \in [0, 8]$. As previously observed, anisotropy induces a modest augmentation in the total mass while exerting only a minimal influence on the stellar radius. (f) Mass function for $\rho_c = 10^{14.5}\,\mathrm{kg/m^3}, \alpha = 0.4, \beta = 0, \kappa \in [0, 8]$. The introduction of electric charge conspicuously elevates the mass curve, whilst also slightly augmenting the stellar radius.

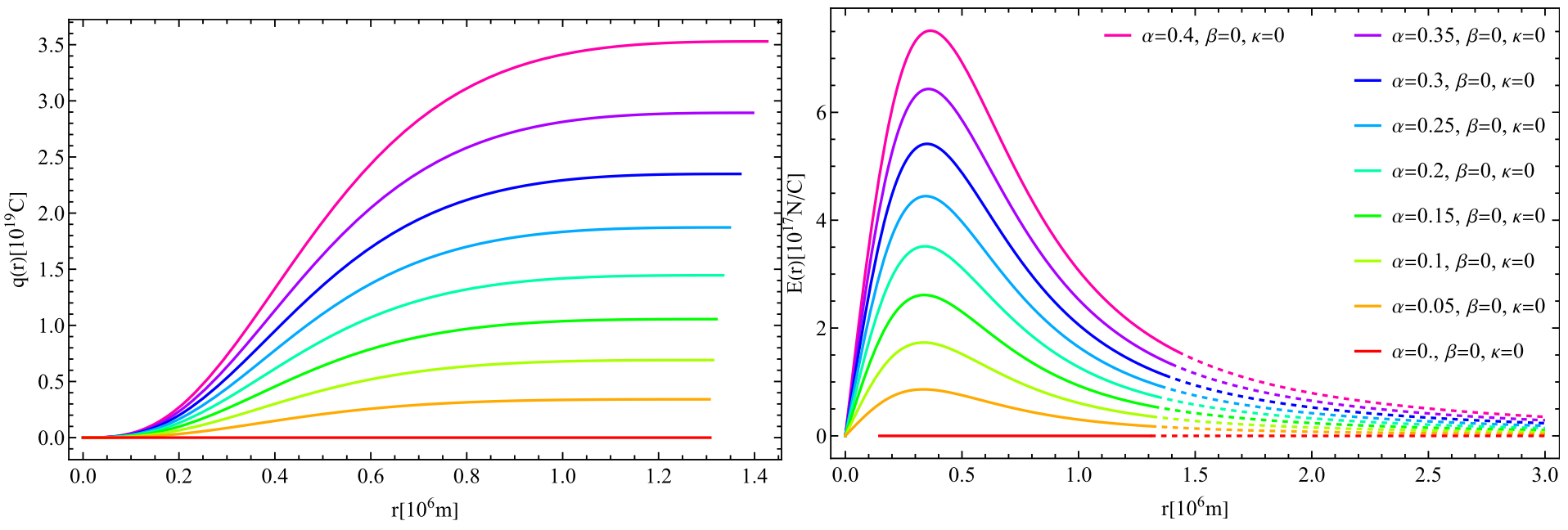

(a)

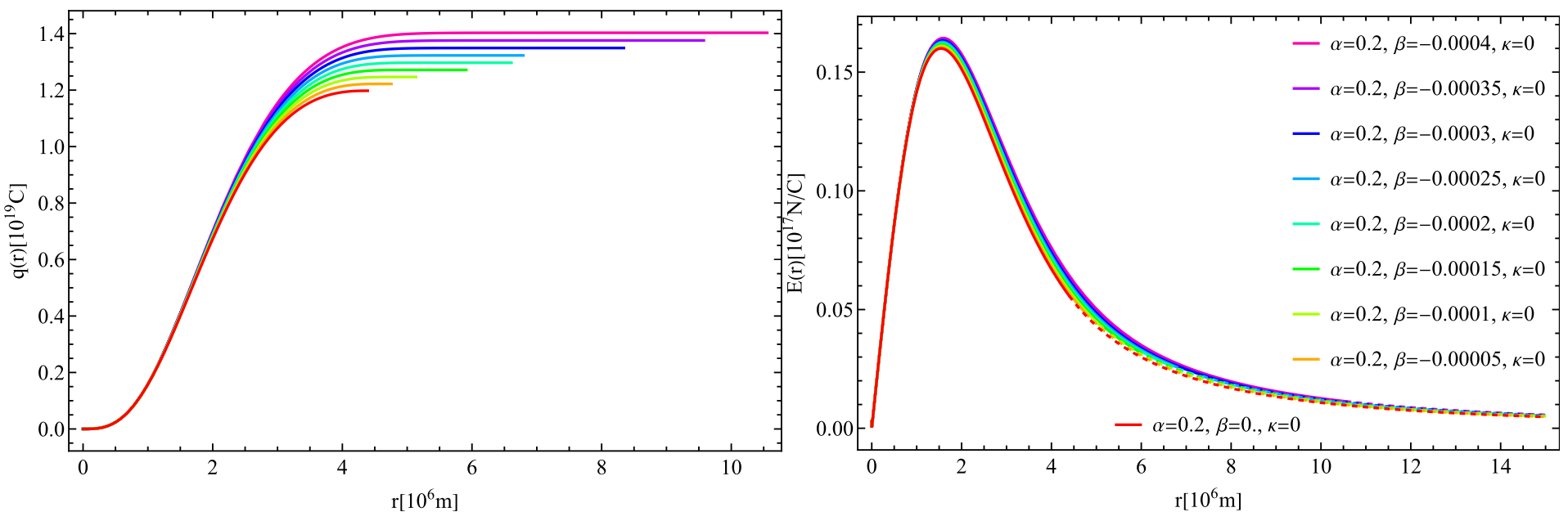

(b)

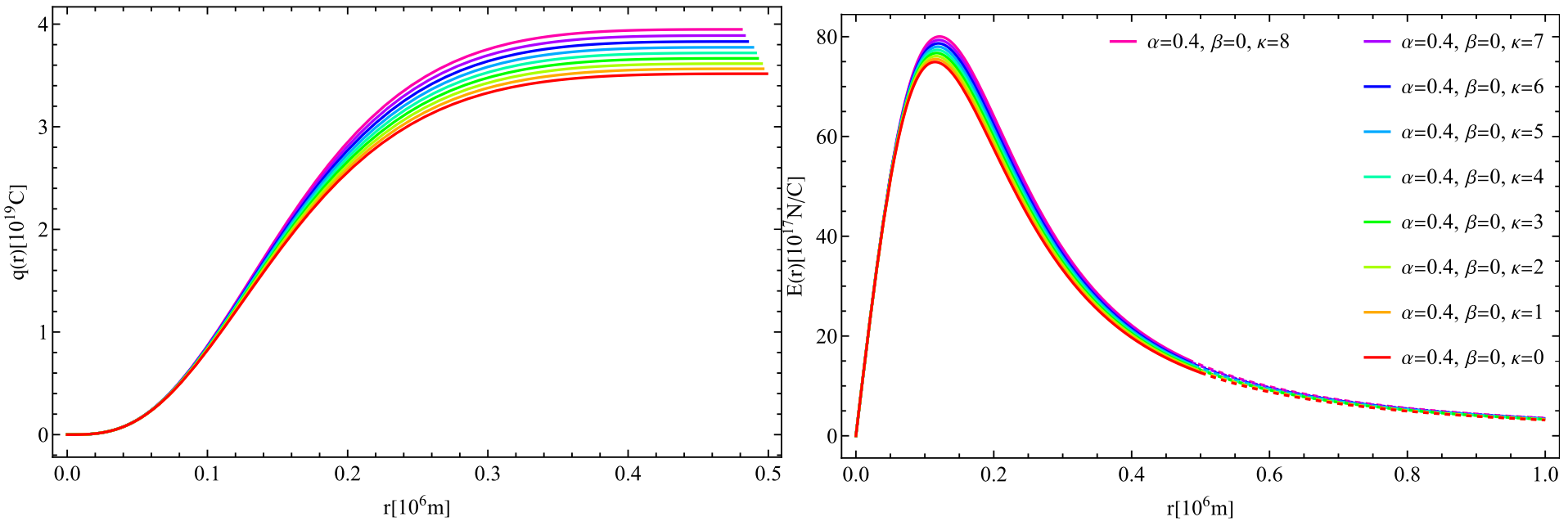

(c)

Fig. 3. Electric charge function $q(r)$ and electric field $E(r)$. We observe that the extremum of the electric field manifests within the stellar interior, a characteristic that diverges from the behavior noted in the literature for quark stars. (a) Electric charge function $q(r)$ and electric field $E(r)$ for $\rho_c = 10^{13}\,\mathrm{kg/m^3}, \alpha \in [0, 0.4], \beta = 0, \kappa = 0$. Naturally, the electric charge and electric field exhibit acute sensitivity to variations in $\alpha$. (b) Electric charge function $q(r)$ and electric field $E(r)$ for $\rho_c = 10^{11}\,\mathrm{kg/m^3}, \alpha = 0.2, \beta \in [-4 \times 10^{-4}, 0], \kappa = 0$. The influence of modified gravity on the charge function aligns with its previously analyzed impact on the mass function, while its effect on the magnitude of the electric field remains relatively insubstantial. (c) Electric charge function $q(r)$ and electric field $E(r)$ for $\rho_c = 10^{14.5}\,\mathrm{kg/m^3}, \alpha = 0.4, \beta = 0, \kappa \in [0, 8]$. Analogous to the effects of modified gravity, the impact of anisotropy on the charge function remains congruent with its influence on the mass function, while its ramifications on the electric field are relatively negligible.

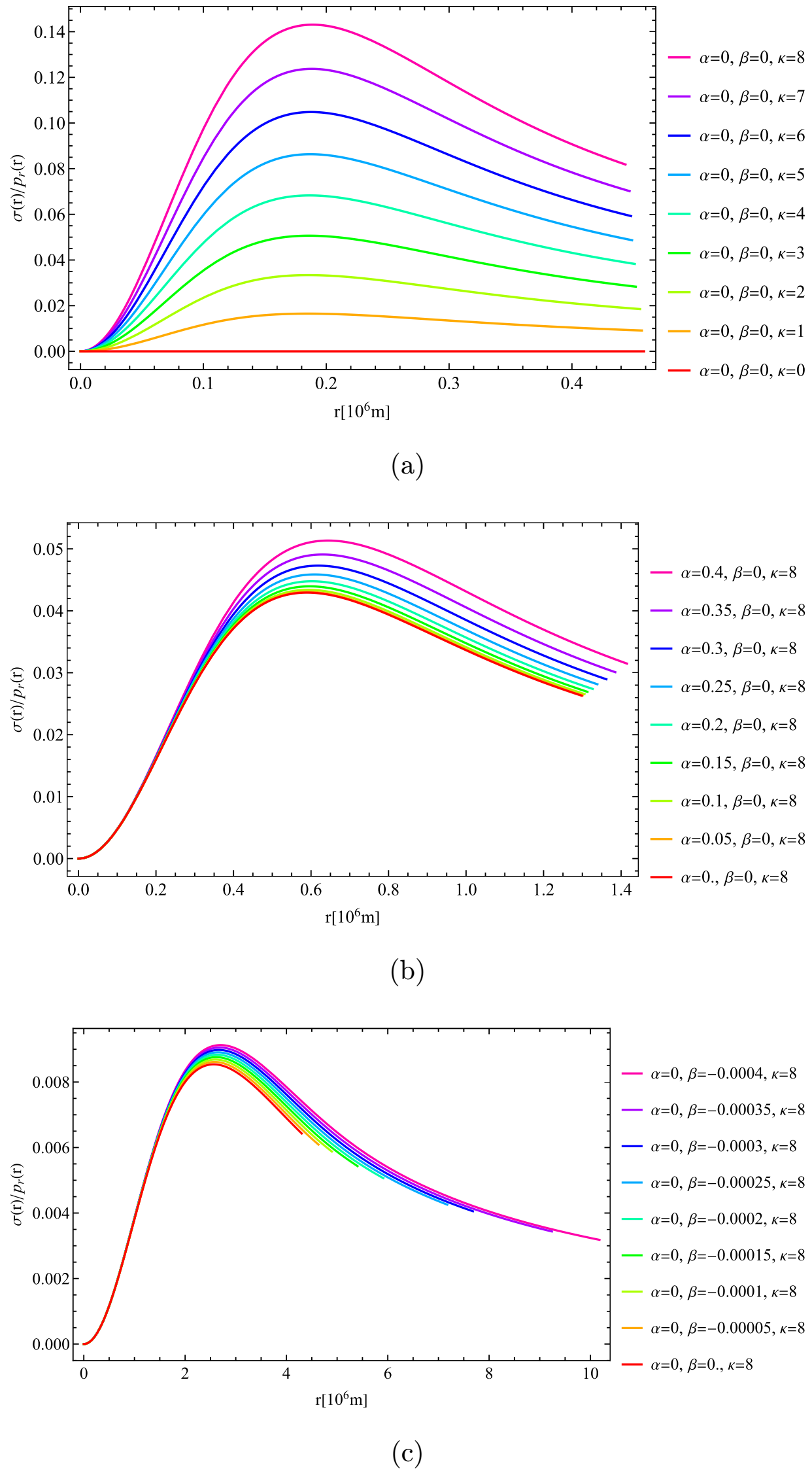

Fig. 4. Measure of relative anisotropy $\sigma/p_r$. In summary, the effects of anisotropy are more pronounced at high densities. It is also noteworthy that the relative maximum of anisotropy likewise occurs within the stellar interior. (a) Relative anisotropy $\sigma/p_r$ for $\rho_c = 10^{14.5}\,\mathrm{kg/m^3}, \alpha = 0, \beta = 0, \kappa \in [0, 8]$. It is self-evident that anisotropy exhibits a relatively uniform dependency on the proportional coefficient $\kappa$. (b) Relative anisotropy $\sigma/p_r$ for $\rho_c = 10^{13}\,\mathrm{kg/m^3}, \alpha \in [0, 0.4], \beta = 0, \kappa = 8$. As the electric charge progressively increases, the relative augmentation of anisotropy accelerates. It can be observed that in the presence of anisotropy, the presence of electric charge similarly expands the stellar radius. (c) Relative anisotropy $\sigma/p_r$ for $\rho_c = 10^{11}\,\mathrm{kg/m^3}, \alpha = 0, \beta \in [-4 \times 10^{-4}, 0], \kappa = 8$. The influence of modified gravity on relative anisotropy is not markedly discernible. As ever, the impact of modified gravity on the stellar radius remains profoundly significant.

stellar surface. The electric field, however, rapidly escalates to its peak value within the stellar interior, subsequently decreasing monotonically, which contrasts distinctly with the results for charged quark stars.[22] The electric field values inside and outside the star are continuous at the stellar surface. Within our model, the electric field remains insensitive to both modified gravity and anisotropy.

Finally, in Fig. 4, we present the measurement of relative anisotropy $\sigma/p_r$. At the selected high central density (Log $\rho_c$[kg/m$^3$] = 14.5), the relative anisotropy reaches up to $\sim 15\%$. Consistent with prior observations, the effect of anisotropy on the stellar structure is more pronounced at high densities. Additionally, both electric charge and modified gravity can enhance relative anisotropy. It is noteworthy that, similar to the electric field distribution, the maximum value of relative anisotropy also occurs within the stellar interior rather than at the surface.

## 5. Conclusions

In this paper, we scrutinize the physical attributes of charged white dwarfs exhibiting anisotropic pressure within the framework of $f(R, T)$ gravity. The gravitational field action in $f(R, T)$ gravity represents a more generalized functional form compared to General Relativity. By treating the white dwarf as a sphere constituted of degenerate electron gas, we employ the Chandrasekhar equation of state. The equation of state, which initially characterizes the relationship between isotropic pressure and density, is construed as radial pressure in this study. In the main text, we allude to other, more intricate equations of state pertaining to white dwarfs. Furthermore, the charge density distribution is postulated to be proportional to the energy density, thereby elucidating the fundamental characteristics of the white dwarf structure.

Upon deriving the modified Tolman–Oppenheimer–Volkoff equations, a series of numerical solutions were obtained under specific central density conditions. By selecting various parameter combinations, we compared the effects of electric charge, modified gravity, and anisotropy on the structure of white dwarfs. The relationships were illustrated in several graphs, including (i) the mass ($M$)–radius ($R_{\rm sur}$) relation and the mass ($M$)-central density (Log $\rho_c$) curve; (ii) the mass function ($m(r)$) as a function of the radial coordinate; (iii) the electric charge distribution ($q(r)$) and the electric field function ($E(r)$); (iv) the relative anisotropy ($\sigma(r)/p_r(r)$) as a function of the radial coordinate. The typical features of each scenario were summarized. The presence of an electric charge increases the radius and total mass of the star under different central densities, with the rate of increase accelerating as the charge density increases. The effects of modified gravity are primarily observed in cases of lower central density, where the stellar radius significantly expands while the total mass only slightly increases. Anisotropy has a greater impact on the high central density side, increasing the total mass of the star while slightly decreasing the stellar radius. Additionally, significant anisotropy (for $\kappa \gtrsim 2$ with $\alpha = 0, \beta = 0$) can even alter the geometric characteristics of the mass–radius relation and the mass-central density curve. Simultaneously, the extremum values of both the electric field and relative anisotropy occur within the stellar interior rather than at the surface. We also pointed out that in this four-dimensional parameter space $(\rho_c, \alpha, \beta, \kappa)$, the region allowing for super-Chandrasekhar behavior is considerably large.

The utility of adopting $f(R,T) = R + 2\beta T$ lies in its computational tractability while ensuring the representation of modified gravitational effects. This model reverts to the canonical General Relativity framework outside the stellar boundary. More intricate models emerge within the general $f(R)$ and $f(R,T)$ theories. The author has endeavored to explore a more elaborate model, $f(R,T) = R + 2\beta T + \gamma RT$, to examine the intricacies of matter–gravity coupling. The analytical derivation of this model parallels the procedures delineated herein, albeit with the emergence of higher-order complexities in numerical solutions. Additionally, other scholars have investigated models such as $f(R,T) = R + 2\beta T + \gamma T^2$ and $f(R,T,R_{ab}T^{ab})$ gravity,[84–86] yielding intriguing results. It is crucial to acknowledge, however, that the gravitational model discussed in this work has been the subject of ongoing debate in recent years, with critical perspectives available in Refs. 87–89.

Once more, it is accentuated that the charge density distribution postulated in this paper is proportional to the matter-energy density, which may not epitomize the optimal white dwarf model. Other scholars have proposed that the charge density distribution of white dwarfs adheres to a Gaussian distribution, centered on the stellar surface,[15,65] and have accordingly derived their respective outcomes. Additionally, as a pivotal issue in the physics of white dwarfs, the radius–mass relationship is subtly influenced by numerous critical factors not addressed in this work, encompassing thermal effects (finite temperature),[29,30,32,90–94] rotation,[27,28,31,48,92,95] nuclear composition,[34,38,47,50–52] magnetization,[23–32] and the like. Further theoretical and observational explorations pertaining to white dwarfs remain a subject of considerable anticipation.

## Appendix A. Nonzero Components of Several Important Tensors

$$\begin{aligned} \mathcal{M}_{00} &= \rho \mathrm{e}^{2\psi}, \quad & \mathcal{M}_{11} &= p_r \mathrm{e}^{2\lambda} \\ \mathcal{M}_{22} &= p_t r^2, \quad & \mathcal{M}_{33} &= p_t r^2 \sin^2\theta \end{aligned} \tag{A.1}$$

$$\begin{aligned} \Theta_{00} &= -(p_t + 2\rho)\mathrm{e}^{2\psi}, \quad & \Theta_{11} &= (p_t - 2p_r)\mathrm{e}^{2\psi} \\ \Theta_{22} &= -p_t r^2, \quad & \Theta_{33} &= -p_t r^2 \sin^2\theta \end{aligned} \tag{A.2}$$

$$\begin{aligned} \mathcal{E}_{00} &= \mathrm{e}^{2\psi} \times q^2/8\pi r^4, \quad & \mathcal{E}_{11} &= -\mathrm{e}^{2\lambda} \times q^2/8\pi r^4 \\ \mathcal{E}_{22} &= r^2 \times q^2/8\pi r^4, \quad & \mathcal{E}_{33} &= r^2 \sin^2\theta \times q^2/8\pi r^4 \end{aligned} \tag{A.3}$$

## Acknowledgments

This paper is not affiliated with any profit-driven organization, including scientific projects or foundations. I would first like to extend my heartfelt gratitude to Juan M. Z. Pretel from *Rio de Janeiro, CBPF*, who, from across the globe, provided invaluable assistance during a challenging period and shared his extensive expertise on the numerical solutions of differential equations. My sincere appreciation also goes to Ruo-Yang Cui from *Xi'an Jiaotong University* (*Notre Dame University* currently), whose unwavering support has been instrumental throughout my academic journey. Furthermore, I am deeply grateful to my family and friends for their unwavering support. While their assistance has been crucial in the completion of this work, any errors remain solely my responsibility.

## ORCID

Zhe Feng https://orcid.org/0000-0002-6377-8544

## References

1. C. L. Bennett, M. Halpern, G. Hinshaw, N. Jarosik, A. Kogut, M. Limon, S. S. Meyer, L. Page, D. N. Spergel, G. S. Tucker, E. Wollack, E. L. Wright, C. Barnes, M. R. Greason, R. S. Hill, E. Komatsu, M. R. Nolta, N. Odegard, H. V. Peiris, L. Verde and J. L. Weiland, *Astrophys. J. Suppl. Ser.* **148**, 1 (2003).
2. D. N. Spergel, L. Verde, H. V. Peiris, E. Komatsu, M. R. Nolta, C. L. Bennett, M. Halpern, G. Hinshaw, N. Jarosik, A. Kogut, M. Limon, S. S. Meyer, L. Page, G. S. Tucker, J. L. Weiland, E. Wollack and E. L. Wright, *Astrophys. J. Suppl. Ser.* **148**, 175 (2003).
3. D. N. Spergel, R. Bean, O. Dore, M. R. Nolta, C. L. Bennett, J. Dunkley, G. Hinshaw, N. Jarosik, E. Komatsu, L. Page, H. V. Peiris, L. Verde, M. Halpern, R. S. Hill, A. Kogut, M. Limon, S. S. Meyer, N. Odegard, G. S. Tucker, J. L. Weiland, E. Wollack and E. L. Wright, *Astrophys. J. Suppl. Ser.* **170**, 377 (2007).
4. D. Glavan and C. Lin, *Phys. Rev. Lett.* **124**, 081301 (2019).
5. S. Shankaranarayanan and J. P. Johnson, *Gen. Relat. Gravit.* **54** (2022), arXiv:2204.06533.
6. T. P. Sotiriou and V. Faraoni, *Rev. Mod. Phys.* **82**, 451 (2010).
7. A. D. Felice and S. Tsujikawa, *Living Rev. Relativ.* **13** (2010), https://doi.org/10.12942/lrr-2010-3.
8. S. Nojiri and S. D. Odintsov, *Phys. Rep.* **505**, 59 (2010).
9. S. Nojiri, S. D. Odintsov and V. K. Oikonomou, *Phys. Rep.* **692**, 1 (2017).
10. A. Starobinsky, *Phys. Lett. B* **91**, 99 (1980).
11. K. Numajiri, T. Katsuragawa and S. Nojiri, *Phys. Lett. B* **826**, 136929 (2021).
12. J. M. Z. Pretel and S. B. Duarte, *Class. Quantum Grav.* **39**, 155003 (2022).
13. T. Harko, F. S. N. Lobo, S. Nojiri and S. D. Odintsov, *Phys. Rev. D* **84**, 024020 (2011).
14. G. A. Carvalho, R. V. Lobato, P. H. R. S. Moraes, J. D. V. Arbail, R. M. M. Jr., E. Otoniel and M. Malheiro, *Eur. Phys. J. C* **77** (2017), arXiv:1706.03596.
15. F. Rocha, G. A. Carvalho, D. Deb and M. Malheiro, *Phys. Rev. D* **101**, 104008 (2019).
16. D. Deb, S. V. Ketov, M. Khlopov and S. Ray, *JCAP* **2019**, 070 (2018).
17. M. Sharif and A. Waseem, *Eur. Phys. J. C* **78** (2018), https://doi.org/10.1140/epjc/s10052-018-6363-2.
18. S. Biswas, D. Shee, B. K. Guha and S. Ray, *Eur. Phys. J. C* **80** (2020), https://doi.org/10.1140/epjc/s10052-020-7725-0.

19. S. Biswas, D. Deb, S. Ray and B. Guha, *Ann. Phys.* **428**, 168429 (2021).
20. P. Rej and P. Bhar, *Astrophys. Space Sci.* **366**, 35 (2021).
21. P. Rej, P. Bhar and M. Govender, *Eur. Phys. J. C* **81**, 316 (2021).
22. J. M. Z. Pretel, T. Tangphati, A. Banerjee and A. Pradhan, *Chin. Phys. C* **46**, 115103 (2022).
23. U. Das and B. Mukhopadhyay, *Phys. Rev. D* **86**, 042001 (2012).
24. U. Das and B. Mukhopadhyay, *Int. J. Mod. Phys. D* **21**, 1242001 (2012).
25. U. Das and B. Mukhopadhyay, *Phys. Rev. D* **110**, 071102 (2013).
26. L. Ferrario, D. de Martino and B. Gaensicke, *Space Sci. Rev.* **191**, 111 (2015).
27. L. Ferrario and D. T. Wickramasinghe, *Mon. Not. R. Astron. Soc.* **356**, 615 (2005).
28. S. Subramanian and B. Mukhopadhyay, *Mon. Not. R. Astron. Soc.* **454**, 752 (2015).
29. S. Yoshida, *Mon. Not. R. Astron. Soc.* **486**, 2982 (2018).
30. J. Peterson, V. Dexheimer, R. Negreiros and B. G. Castanheira, *Astrophys. J.* **921**, 1 (2021).
31. I. Caiazzo, K. B. Burdge, J. Fuller, J. Heyl, S. R. Kulkarni, T. A. Prince, H. B. Richer, J. Schwab, I. Andreoni, E. C. Bellm, A. Drake, D. A. Duev, M. J. Graham, G. Helou, A. A. Mahabal, F. J. Masci, R. Smith and M. T. Soumagnac, *Nature* **595**, 39 (2021).
32. R. Sahoo, T. P. Anant and S. Mukhopadhyay, *J. Astrophys. Astron.* **45** (2024), https://doi.org/10.1007/s12036-024-10015-x.
33. K. Maguire, M. Eracleous, P. G. Jonker, M. MacLeod and S. Rosswog, *Space Sci. Rev.* **216** (2020), arXiv:2004.00146.
34. A. Bobrick, Y. Zenati, H. B. Perets, M. B. Davies and R. Church, *Mon. Not. R. Astron. Soc.* **510**, 3758 (2021).
35. J.-H. Chen, R.-F. Shen and S.-F. Liu, *Astrophys. J.* **947**, 32 (2022).
36. G. Panotopoulos and I. Lopes, *Int. J. Mod. Phys. D* **29**, 2050058 (2020).
37. C. J. Horowitz, *Phys. Rev. D* **102**, 083031 (2020).
38. N. F. Bell, G. Busoni, M. E. Ramirez-Quezada, S. Robles and M. Virgato, *JCAP* **2021**, 083 (2021).
39. H. Steigerwald, V. Marra and S. Profumo, *Phys. Rev. D* **105**, 083507 (2022).
40. M. A. Hollands, P.-E. Tremblay, B. T. Gnsicke, N. P. Gentile-Fusillo and S. Toonen, *Mon. Not. R. Astron. Soc.* **480**, 3942 (2018).
41. J. McCleery, P.-E. Tremblay, N. P. Gentile Fusillo, M. A. Hollands, B. T. Gnsicke, P. Izquierdo, S. Toonen, T. Cunningham and A. Rebassa-Mansergas, *Mon. Not. R. Astron. Soc.* **499**, 1890 (2020).
42. M. Kilic, P. Bergeron, S. Blouin and A. Bdard, *Mon. Not. R. Astron. Soc.* **503**, 5397 (2021).
43. N. P. Gentile Fusillo, P.-E. Tremblay, E. Cukanovaite, A. Vorontseva, R. Lallement, M. Hollands, B. T. Gnsicke, K. B. Burdge, J. McCleery and S. Jordan, *Mon. Not. R. Astron. Soc.* **508**, 3877 (2021).
44. P.-E. Tremblay, A. Bdard, M. W. O'Brien, J. Munday, A. K. Elms, N. P. Gentillo Fusillo and S. Sahu, *New Astron. Rev.* **99**, 101705 (2024).
45. V. Korol, N. Hallakoun, S. Toonen and N. Karnesis, *Mon. Not. R. Astron. Soc.* **511**, 5936 (2021).
46. M. Kilic, A. Kosakowski, A. G. Moss, P. Bergeron and A. A. Conly, *Astrophys. J. Lett.* **923**, L6 (2021).
47. M. Rotondo, J. A. Rueda, R. Ruffini and S. S. Xue, *Phys. Rev. D* **84**, 084007 (2010).
48. K. Boshkayev, J. A. Rueda, R. Ruffini and I. Siutsou, *Astrophys. J.* **762**, 117 (2012).
49. K. Boshkayev, H. Quevedo and B. Zhami, *Mon. Not. R. Astron. Soc.* **464**, 4349 (2016).
50. P. Bergeron, P. Dufour, G. Fontaine, S. Coutu, S. Blouin, C. Genest-Beaulieu, A. Bdard and B. Rolland, *Astrophys. J.* **876**, 67 (2019).

51. A. H. Crsico, L. G. Althaus, M. M. M. Bertolami and S. O. Kepler, *Astron. Astrophys. Rev.* **27** (2019), arXiv:1907.00115.
52. L. G. Althaus, A. H. Crsico, M. E. Camisassa, S. Torres, P. Gil-Pons, A. Rebassa-Mansergas and R. Raddi, *Mon. Not. R. Astron. Soc.* **523**, 4492 (2023).
53. M. F. Sousa, J. G. Coelho and J. C. N. de Araujo, *Mon. Not. R. Astron. Soc.* **492**, 5949 (2020).
54. M. Georgousi, N. Karnesis, V. Korol, M. Pieroni and N. Stergioulas, *Mon. Not. R. Astron. Soc.* **519**, 2552 (2022).
55. C.-Q. Ye, J.-H. Chen, J.-D. Zhang, H.-M. Fan and Y.-M. Hu, *Mon. Not. R. Astron. Soc.* **527**, 2756 (2023).
56. D.-H. Wen, H.-L. Liu and X.-D. Zhang, *Chin. Phys. B* **23**, 089501 (2014).
57. G. A. Carvalho, R. M. Marinho and M. Malheiro, *Gen. Relat. Gravit.* **50** (2017), arXiv:1709.01635.
58. L. Perot and N. Chamel, *Phys. Rev. D* **106**, 023012 (2022).
59. S. Chandrasekhar, *Astrophys. J.* **74**, 81 (1931).
60. D. A. Howell, M. Sullivan, P. E. Nugent, R. S. Ellis, A. J. Conley, D. L. Borgne, R. G. Carlberg, J. Guy, D. Balam, S. Basa, D. Fouchez, I. M. Hook, E. Y. Hsiao, J. D. Neill, R. Pain, K. M. Perrett and C. J. Pritchet, *Nature* **443**, 308 (2006).
61. R. A. Scalzo, G. Aldering, P. Antilogus, C. Aragon, S. Bailey, C. Baltay, S. Bongard, C. Buton, M. Childress, N. Chotard, Y. Copin, H. K. Fakhouri, A. Gal-Yam, E. Gangler, S. Hoyer, M. Kasliwal, S. Loken, P. Nugent, R. Pain, E. Pcontal, R. Pereira, S. Perlmutter, D. Rabinowitz, A. Rau, G. Rigaudier, K. Runge, G. Smadja, C. Tao, R. C. Thomas, B. Weaver and C. Wu, *Astrophys. J.* **713**, 1073 (2010).
62. S. Chandrasekhar, *Mon. Not. R. Astron. Soc.* **95**, 207 (1935).
63. H. Liu, X. Zhang and D. Wen, *Phys. Rev. D* **89**, 104043 (2014).
64. A. V. Astashenok, S. D. Odintsov and V. K. Oikonomou, *Phys. Rev. D* **106**, 124010 (2022).
65. G. A. Carvalho, J. D. V. Arbail, R. M. Marinho and M. Malheiro, *Eur. Phys. J. C* **78** (2018), https://doi.org/10.1140/epjc/s10052-018-5901-2.
66. A. S. Eddington, *The Internal Constitution of the Stars* (Cambridge Univ. Press, 1988).
67. S. Ray, A. L. Espindola, M. Malheiro, J. P. S. Lemos and V. T. Zanchin, *Phys. Rev. D* **68**, 084004 (2003).
68. J. P. S. Lemos, F. J. Lopes, G. Quinta and V. T. Zanchin, *Eur. Phys. J. C* **75**, 76 (2014).
69. J. D. V. Arbail and V. T. Zanchin, *Phys. Rev. D* **97**, 104045 (2017).
70. G. Panotopoulos, A. Pradhan, T. Tangphati and A. Banerjee, *Chin. J. Phys.* **77**, 2106 (2022).
71. R. P. Negreiros, F. Weber, M. Malheiro and V. Usov, *Phys. Rev. D* **80**, 083006 (2009).
72. C. Zhang, M. Gammon and R. B. Mann, *Phys. Rev. D* **104**, 123007 (2021).
73. J. M. Pretel, S. E. Jors, R. R. Reis and J. D. Arbail, *JCAP* **2021**, 064 (2021).
74. M. Sharif and A. Siddiqa, *Eur. Phys. J. Plus* **133** (2018), https://doi.org/10.1140/epjp/i2018-12060-8.
75. S. Maurya and F. Tello-Ortiz, *Phys. Dark Univ.* **27**, 100442 (2020).
76. D. Deb, F. Rahaman, S. Ray and B. K. Guha, *JCAP* **2018**, 044 (2017).
77. T. Tangphati, I. Karar, A. Pradhan and A. Banerjee, *Eur. Phys. J. C* **82** (2022), https://doi.org/10.1140/epjc/s10052-022-10024-6.
78. V. Faraoni, *Phys. Rev. D* **80**, 124040 (2009).
79. T. Harko, *Phys. Rev. D* **81**, 044021 (2010).
80. E. E. Salpeter, *Astrophys. J.* **134**, 669 (1961).
81. T. Hamada and E. E. Salpeter, *Astrophys. J.* **134**, 683 (1961).

82. A. DeBenedictis, D. Horvat, S. Ilijic, S. Kloster and K. S. Viswanathan, *Class. Quantum Grav.* **23**, 2303 (2005).
83. D. Horvat, S. Ilijic and A. Marunovic, *Class. Quantum Grav.* **28**, 025009 (2010).
84. M. Sharif and A. Waseem, *Can. J. Phys.* **94**, 1024 (2016).
85. M. Sharif and A. Waseem, *Int. J. Mod. Phys. D* **27**, 1950007 (2018).
86. M. Sharif and T. Naseer, *Chin. J. Phys.* **73**, 179 (2021).
87. S. B. Fisher and E. D. Carlson, *Phys. Rev. D* **100**, 064059 (2019).
88. T. Harko and P. H. Moraes, *Phys. Rev. D* **101**, 108501 (2020).
89. S. B. Fisher and E. D. Carlson, *Phys. Rev. D* **101**, 108502 (2020).
90. K. Boshkayev, J. A. Rueda, B. Zhami, Z. Kalymova and G. Balgymbekov, *Int. J. Mod. Phys. Conf. Ser.* **41**, 1660129 (2015).
91. R. Fantoni, *J. Stat. Mech.* **2017**, 113101 (2017).
92. K. Boshkayev, *Astron. Rep.* **62**, 847 (2018).
93. S. P. Nunes, J. D. V. Arbail and M. Malheiro, *Astrophys. J.* **921**, 138 (2021).
94. J. D. M. Tuacao, A. G. Abac and R. E. S. Otadoy, *Int. J. Mod. Phys. D* **32** (2022), arXiv:2210.09886.
95. P. Kumar and D. M. Townsley, *Astrophys. J.* **951**, 122 (2023).